\documentstyle[pre,aps,preprint,epsf]{revtex}

\begin{document}

\title{Scaling analysis of a divergent prefactor in the metastable
lifetime of a square-lattice Ising ferromagnet at low temperatures}
\draft 
\author{Kyungwha ${\mathrm Park^{1,\ast}}$ 
\and M.~A.\ ${\mathrm Novotny^{2,\dag}}$ 
\and P.~A.\ ${\mathrm Rikvold^{1,3,\ddag}}$}
\address{${\mathrm ^{1}}$School of Computational Science and 
Information Technology, Florida State University, Tallahassee, 
Florida 32306 \\
${\mathrm ^{2}}$Department of Physics and Astronomy, Mississippi 
State University, Mississippi State, Mississippi 39762 \\
${\mathrm ^{3}}$Center for Materials Research and 
Technology and Department of Physics, 
Florida State University, Tallahassee, Florida 32306}
\date{\today}
\maketitle

\newpage

\begin{abstract}
We examine a square-lattice nearest-neighbor Ising quantum ferromagnet 
coupled to $d$-dimensional phonon baths. Using the density-matrix equation,
we calculate the transition rates between configurations, which determines
the specific dynamic. Applying the calculated stochastic dynamic 
in Monte Carlo simulations, we measure 
the lifetimes of the metastable state. As the magnetic field approaches
$|H|/J=2$ at low temperatures, the lifetime prefactor diverges
because the transition rates between certain configurations 
approaches zero under these conditions. Near $|H|/J=2$ and zero temperature, 
the divergent prefactor shows scaling behavior as a function of the 
field, temperature, and the dimension of the phonon baths. 
With proper scaling, the simulation data at different 
temperatures and for different dimensions of the baths collapse well 
onto two master curves, one for $|H|/J>2$ and one for $|H|/J<2$.
\end{abstract}


\section{Introduction}
Compared to the exponentials that occur in the expressions for
particular physical quantities, the associated
prefactors are often assumed to be too unremarkable and uninteresting 
to examine, so that the study of
prefactors has been ignored in many cases. However, this is not always 
true. In this paper, we explore interesting behavior shown by the 
prefactors of the lifetime of the metastable state of a two-dimensional
nearest-neighbor Ising ferromagnet interacting with phonon baths. 
Metastability occurs often in many different systems, ranging from 
supercooled fluids and vapors\cite{ABRA74,OXTO92} to spin 
systems\cite{RIKV94} and quantum liquids\cite{LEGG84}. In those systems
some type of weak noise (for example, thermal noise) can drive the system
from the metastable 
state to the stable state across a saddle point at an extremely small rate. 
Therefore, the average waiting time before escape from a metastable
state is usually extremely long.

Recently, we examined a square-lattice Ising quantum ferromagnet with
a phonon (i.e., bosonic) bath attached to each spin, in the presence of a 
longitudinal magnetic field\cite{PARK02-1,PARK02-2}. 
The time evolution of the spin system is determined by the linear 
coupling of the system with the phonon baths, 
and the dynamic (defined by the transition rates between different spin 
configurations) 
is calculated using the quantum-mechanical density-matrix equation. The
resulting dynamic is different from the Glauber dynamic\cite{GLAU63}, which 
can be derived in a similar way 
from coupling with fermionic baths\cite{MART77}. 
In the present paper, the calculated transition rates are 
applied in dynamic Monte Carlo simulations in order to measure the 
lifetimes of the metastable state. To measure the lifetimes, the initial 
configuration is set to be all spins up, and a magnetic field which 
favors all spins down is applied. Then the number of spin-flip attempts 
is measured until the system magnetization reaches zero. 
At a particular value of the magnetic field, $|H|/J=2$ where $J$ is the
nearest-neighbor exchange coupling constant, the present 
dynamic does not allow transitions between certain configurations, 
which causes the lifetime to diverge as the temperature approaches 
zero and $|H|/J \rightarrow 2^{\pm}$. 
Since the energy barrier between the stable and metastable states 
is finite, this means that the prefactor of the lifetime
must diverge as $|H|/J \rightarrow 2^{\pm}$. If the Glauber dynamic is applied 
to the system instead, then the lifetime prefactor has a finite, 
field-independent value for $|H|/J>2$, which is different from that 
for $|H|/J<2$\cite{NOVO97,NOVO02}.
Consequently, the divergence of the prefactor is due to the specific 
dynamic imposed on the spin system. From the dynamic Monte Carlo 
simulations, we demonstrate that this divergent prefactor reveals 
a scaling behavior which is determined by $T^d$ and the 
ratio $(|H|-2J)/(k_B T)$, 
where $T$ is the absolute temperature, $k_B$ is Boltzmann's constant,
and $d$ is the dimension of the phonon baths.

Quite recently, Maier and Stein investigated the effects of weak 
noise on the magnetization reversal rate in an overdamped one-dimensional 
classical Ginzburg-Landau model of finite length\cite{MAIE01}. 
There is no applied field in their model, so thermal fluctuations drive
the system from one stable configuration (magnetization near +1) to another 
(magnetization near $-1$) through a saddle point.
They analytically calculated the prefactor of the magnetization 
reversal rate, and found that the rate prefactor diverges as the system 
size $L$ reaches $\pi$ or $2\pi$, depending on the boundary conditions. 
A similar prefactor divergence occurs in 
a two-dimensional nonequilibrium model\cite{MAIE96}.
Shneidman and Nita\cite{SHNE02} studied a metastable lattice-gas model with
nearest-neighbor interactions and continuous-time Metropolis dynamics. 
They calculated 
analytically the prefactor of the lifetime of the metastable state 
beyond the zero-temperature limit. The lifetime prefactor exhibited 
distinctive peaks as a function of field, which disappear at zero
temperature.
 
The remainder of this paper is organized as follows. In Sec.~II 
we describe our dynamic quantum model, and in Sec.~III
we show how to calculate the prefactor analytically and numerically.
In Sec.~IV we present our scaling analysis of the prefactor, and
in Sec.~V we present a discussion and our conclusions.

\section{Dynamic Quantum Model}

To derive the classical dynamic (i.e., transition rates) 
from the quantum system,
we use the Hamiltonian: 
\begin{eqnarray}
{\mathcal H}&=&{\mathcal H}_{\mathrm{sp}}+{\mathcal H}_{\mathrm{ph}}
+{\mathcal H}_{\mathrm {sp-ph}} \;, \\
{\mathcal H}_{\mathrm {sp}} &=& -J \sum_{\langle i,j \rangle} 
\sigma_i^z \sigma_j^z -H \sum_{i} \sigma_i^z \;, 
\label{eq:hspin} \\
{\mathcal H}_{\mathrm {ph}}&=& 
\sum_{q} \hbar \omega_{q} c^{\dagger}_{q} c_{q} \;, \\
{\mathcal H}_{\mathrm {sp-ph}} &=& \lambda  \sum_{j,q}
 \sqrt{\!\frac{\hbar}{2NM\omega_{q}}} (iq~ \sigma_j^x) 
(c^{\dagger}_{q} - c_{q}) e^{i q R_j} \;, 
\end{eqnarray}
where ${\mathcal H}_{\mathrm {sp}}$ is the spin Hamiltonian,
${\mathcal H}_{\mathrm {ph}}$ is the phonon Hamiltonian, and 
${\mathcal H}_{\mathrm {sp-ph}}$\cite{HART96} is the Hamiltonian 
describing the simple linear spin-phonon coupling.
The first sum in Eq.~(\ref{eq:hspin}) runs over nearest-neighbor 
sites only on a two-dimensional square lattice, $J(>0)$ is the ferromagnetic 
exchange coupling constant, $\sigma_j^z$ are the $z$ components of Pauli spin 
operators attached to lattice site $j$ (in our notation, their eigenvalues
are $\pm1$), and $H$ is a longitudinal magnetic 
field. The index $q$ is the wave vector of a phonon mode, $\omega_{q}$ is 
the angular frequency of the phonon mode with wave vector $q$, and 
$c^{\dagger}_{q}$ and $c_{q}$ are the corresponding creation and annihilation 
operators. The constant $\lambda$ (its dimension is energy) is the coupling 
strength between the spin system and the phonon heat bath, 
$N$ is the number of unit cells in the system, $M$ is the mass per unit cell, 
and $R_j$ is the position of site $j$.
Details of this model were presented in Refs.\cite{PARK02-1,PARK02-2},
and thus here we only briefly sketch the main ideas.

\begin{figure}
\begin{center}
\leavevmode
\epsfxsize=8.5cm
\epsfysize=7.5cm
\epsfbox{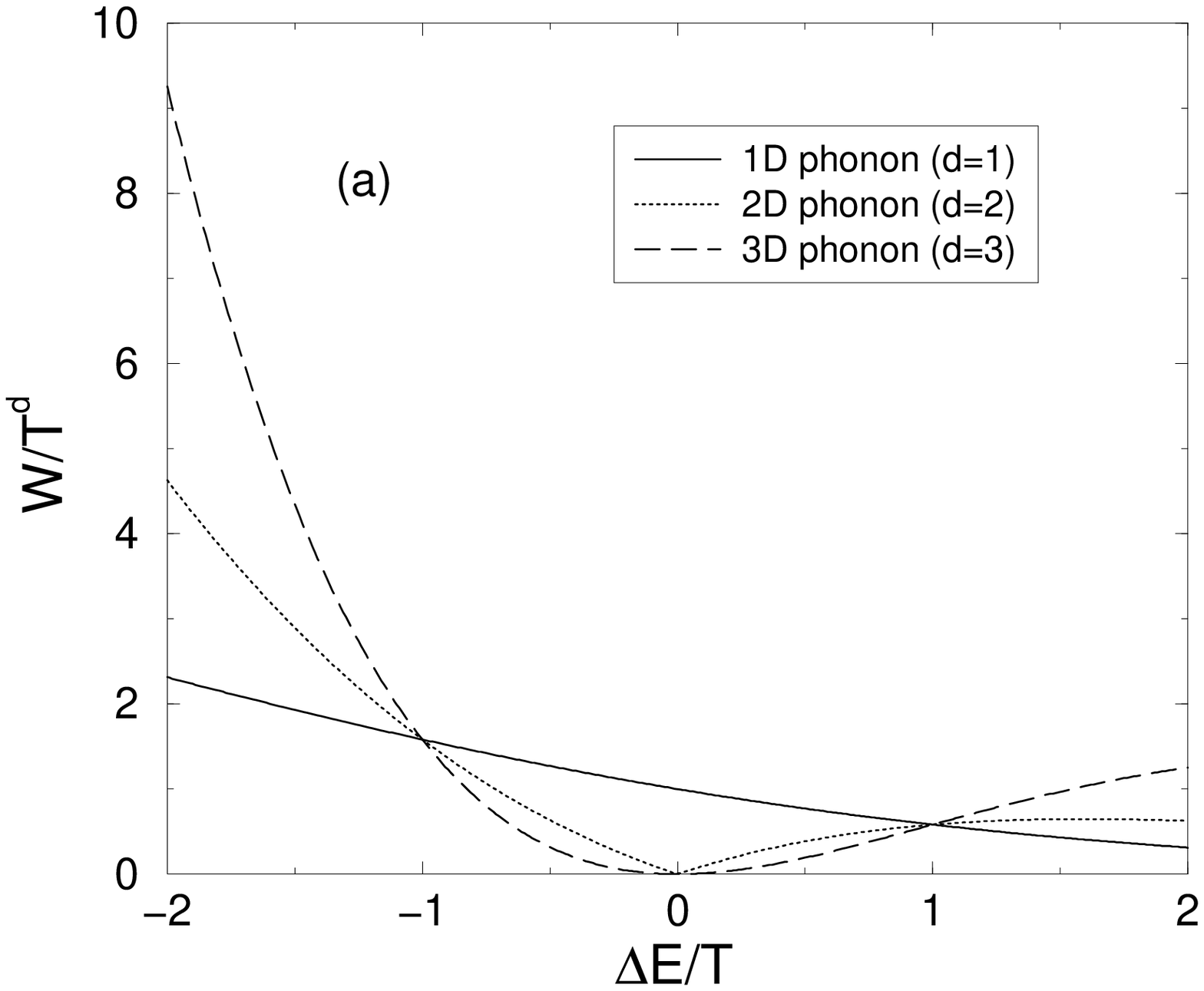}
\epsfxsize=8.5cm
\epsfysize=7.5cm
\epsfbox{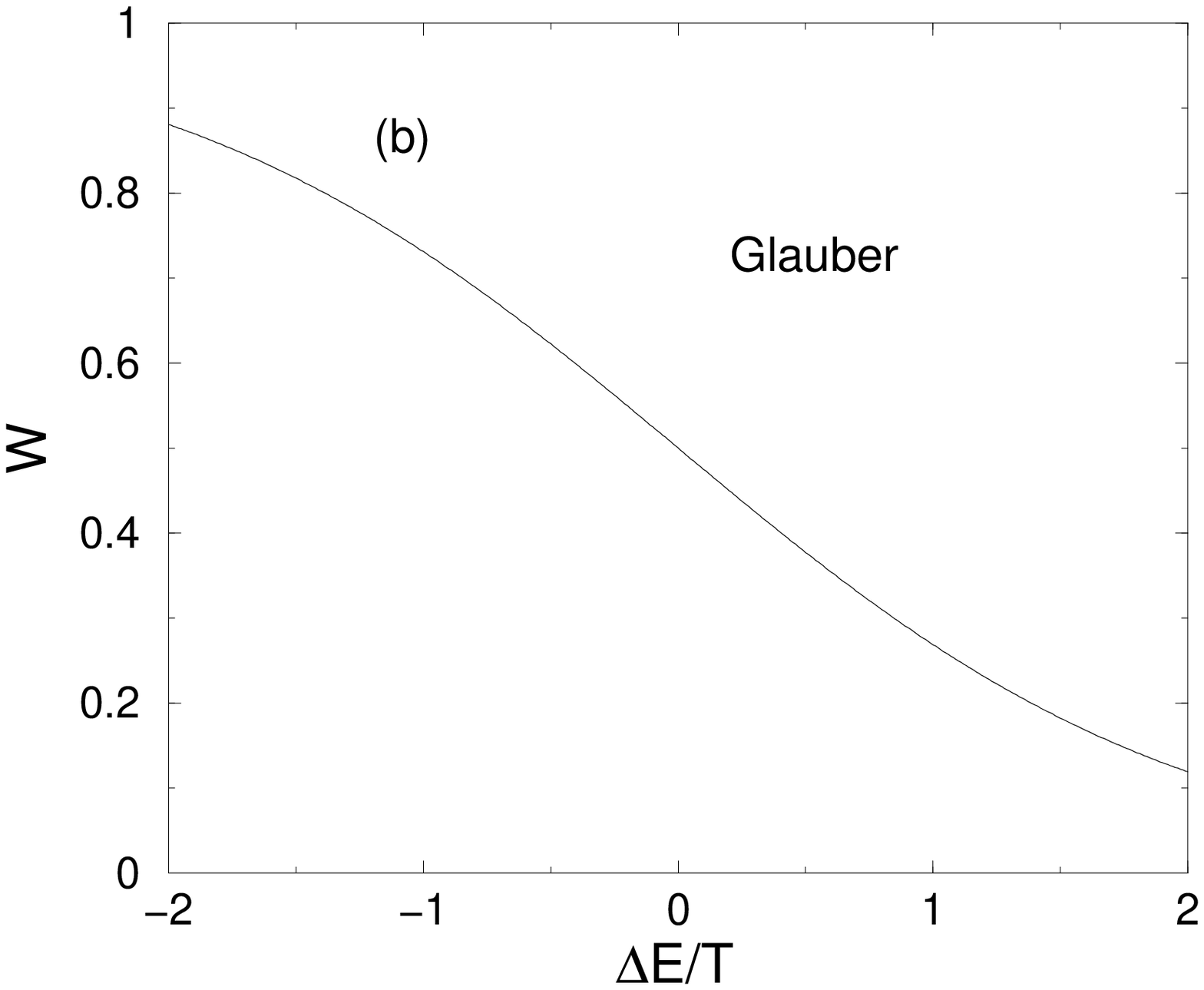}
\caption{(a) Transition rate $W$ divided by $T^d$ vs the energy difference
$\Delta E$ scaled by $T$ for the $d$-dimensional phonon dynamics.
(We use units such as $k_B = 1$.) 
Here we ignore the proportionality constant in $W$, setting
$\lambda^2/\Theta \eta \hbar^{d+1} c^{d+2}=1$.
(b) Transition rate $W$ vs $\Delta E/T$ for the Glauber dynamic.}
\label{fig:WvsE}
\end{center}
\end{figure}

With the given spin Hamiltonian, the dynamic
is determined by the generalized master equation \cite{BLUM96,LEUE00}:
\begin{eqnarray}
\frac{d\rho(t)_{m^{\prime} m}}{dt}&=&\frac{i}{\hbar}
[\rho(t),{\mathcal H}_{\mathrm {sp}}]_{m^{\prime} m} 
+ \delta_{m^{\prime} m} \sum_{n \neq m} \rho(t)_{nn} W_{mn} \nonumber \\
& &- \gamma_{m^{\prime} m} \rho(t)_{m^{\prime} m} \;, 
\nonumber \\
\gamma_{m^{\prime} m}&=&\frac{W_m+W_{m^{\prime}}}{2}\;, \; \; \;
W_m = \sum_{k \neq m} W_{km} \; ,
\end{eqnarray}
where $\rho(t)$ is the time dependent density matrix of the 
spin system, $m^{\prime}$, $n$, $k$, and $m$ are eigenstates of 
${\mathcal H}_{\mathrm{sp}}$, 
$\rho(t)_{m^{\prime} m}$$=$$\langle m^{\prime}| \rho(t) | m \rangle$,
and $W_{km}$ is the transition rate from the $m$th to the $k$th eigenstate.
Assuming that the correlation time of the heat bath is much smaller
than the times of interest, we integrate over all degrees of freedom of 
the bath. Then the transition rate from the $m$th to the $k$th eigenstate 
of ${\mathcal H}_{\mathrm{sp}}$ becomes\cite{PARK02-1,PARK02-2}
\begin{eqnarray}
W_{k m} &=& \frac{2\pi}{\hbar} \! \sum_{q,n_{q}} \!
\left|{ \langle n_{q} + 1,
k \: |{\mathcal H}_{\mathrm {sp-ph}}|~n_{q},m \rangle }
\right|^2 \langle n_{q} |\rho_{\mathrm{ph}}|n_{q} \rangle
\: \delta(E_m - E_k - \hbar \omega_{q}) \;,
\end{eqnarray}
where $E_m$ and $E_k$ are the energy eigenvalues of 
${\mathcal H}_{\mathrm {sp}}$ and $E_m > E_k$.  Here
$n_q$ is the average occupation number of the phonon mode with
$q$, and $\rho_{\mathrm{ph}}$ is the density matrix of the phonon
bath.  We can calculate the transition rate when $E_m < E_k$ similarly.
Eventually we obtain
\begin{eqnarray}
W_{k m} &=& \frac{\lambda^2}{\Theta \eta \hbar^{d+1} c^{d+2}} \left|{
\frac{(E_k-E_m)^d}{e^{\beta(E_k-E_m)}-1} }\right|~,
\label{eq:wkl}
\end{eqnarray}
where $d$ is the dimension of the heat bath, 
$\Theta=2$ ($2\pi$) for $d=1,2$ (3), 
$\eta=M/a^d$ where $a$ is the lattice constant, $c$ is the sound
velocity, and $\beta=1/k_B T$. 
The two major differences from the Glauber dynamic are
the energy difference term in the numerator and the negative
sign in the denominator. In the limit $T\rightarrow 0$, 
the transition rates vanish when $E_m=E_k$ (this can occur for
$|H|=2J, 4J$). Despite their unusual form, these transition rates 
satisfy detailed balance. The transition rates for different $d$ 
scaled by $(k_B T)^d$ are shown in Fig.~\ref{fig:WvsE}(a) 
as functions of $\beta(E_k-E_m)$.
This can be compared with the transition rate for the Glauber dynamic,
shown in Fig.~\ref{fig:WvsE}(b).

\section{Prefactor: Theory and Simulation}

We apply the calculated transition rates for the $d$-dimensional
phonon dynamic in Monte Carlo simulations and measure the
lifetimes of the metastable state. [Here we consider $H<0$ only
in Eq.~(\ref{eq:hspin}).] First, we summarize
theoretical predictions for the lifetime.
As $T \rightarrow 0$, the exact prediction \cite{JORD91,BOVI01} for the 
energy barrier between the metastable and stable states is given by
\begin{equation}
\Gamma(H,J)=8J\ell_c - 2 |H| (\ell_c^2 - \ell_c + 1) \;,
\label{eq:taulowT}
\end{equation}
where the linear critical droplet size in units of the lattice constant is 
$\ell_c=\lceil 2J/|H| \rceil$, and
$\lceil x \rceil$ denotes the smallest integer not less than
$x$. For example, for $2J<|H|<4J$, $\ell_c=1$, and
for $J<|H|<2J$, $\ell_c=2$. (At $|H|=2$, the critical droplet
size changes.) 
For $\ell_c > 1$, the critical droplet is a cluster of overturned 
spins, which is 
an $\ell_c \times (\ell_c -1)$ rectangle with one additional overturned
spin on one of its long sides. For $\ell_c=1$, a single overturned spin
is the critical droplet.
This formula is valid only when $2J/|H|$ is not an integer and $|H|<4J$.
At low temperatures the mean lifetime is the inverse of the
probability of escaping from the metastable well, which can be
written as\cite{JORD91}
\begin{equation}
\langle \tau \rangle = A(H,T) \exp(\Gamma/T) \:,
\label{eq:lifetime}
\end{equation}
where $A(H,T)$ is the prefactor. In this equation and hereafter
we set $k_B=1$. 
We measure $\langle \tau \rangle$
in units of Monte Carlo spin-flip attempts (MCS). 
In our case, a unit Monte Carlo step is defined to be 
the inverse of the coefficient of the transition rate 
[Eq.~(\ref{eq:wkl})] multiplied by $J^d$:
$\Theta \eta \hbar^{d+1} c^{d+2}/(\lambda^2 J^d)$.
(For the single-droplet decay mode considered here, 
$\langle \tau \rangle$ is independent of the system
size if measured in MCS\cite{RIKV94,RIKV94-2}.)
Hereafter, the field $H$ and temperature $T$ are given in units of $J$,
and we set $J=1$. 

The lifetime can be calculated analytically in terms of the shrinking 
and growing probabilities of a droplet, using a rejection-free 
technique\cite{BORT75} and estimating the first-passage time until 
the system contains a critical droplet and overcome the 
barrier\cite{NOVO97,NOVO01}. Then from Eqs.~(\ref{eq:wkl}) 
and~(\ref{eq:lifetime}),
the prefactor $A$ can be calculated\cite{PARK02-2} as a function of the field
with the $d$-dimensional phonon dynamic, using absorbing Markov 
chains (AMC)\cite{NOVO97,NOVO01} in the limit $T \rightarrow 0$.
For $2 < |H| < 4$,
\begin{equation}
A(H,T \rightarrow 0,d)
=\frac{4(2|H|-4)^d + (8-2|H|)^d}{4(2|H|-4)^d (8-2|H|)^d}\;.
\label{eq:a1}
\end{equation}
For $1 < |H| < 2$,
\begin{equation}
A(H,T \rightarrow 0,d)=\frac{|H|^d + 2(2-|H|)^d}{2^{d+3}|H|^d (2-|H|)^d}\;.
\label{eq:a2}
\end{equation}
Near $|H|=2$, both below and above, we find from Eqs.~(\ref{eq:a1}) 
and (\ref{eq:a2}) that for $|H|$ fixed,
\begin{eqnarray}
\lim_{T \rightarrow 0,\; |H| \neq 2} A(H,T,d) 
&=& \frac{1}{C_1 \: ||H|-2|^d} \;, 
\label{eq:T0}
\end{eqnarray}
where $C_1=2^{d+2}$ for $|H|>2$ and $C_1=2^{d+3}$ for $|H|<2$.
On the other hand, as $|H| \rightarrow 2^{\pm}$ at nonzero $T$, 
with the $d$-dimensional phonon dynamic, the transition rate, 
$W_{km}$ [Eq.~(\ref{eq:wkl})], approaches $T||H|-2|^{d-1}$ times 
a constant. Since the transition rate is inversely
proportional to the prefactor, we find that for $T$ fixed,
\begin{equation}
\lim_{|H| \rightarrow 2^{\pm},\; T > 0} A(H,T,d)
= \frac{C_2}{T ||H|-2|^{d-1}} \;,
\label{eq:h0}
\end{equation}
where the constant $C_2$ may depend on $d$. The exact value of $C_2$ 
could not be obtained analytically, so we examine it numerically (at the
end of Sec.~IV).
As $|H| \rightarrow 2^{\pm}$, for $d=1$ the prefactor approaches $C_2/T$
so that there is no field dependence, but 
for $d=2$ and 3 the prefactor diverges as $C_2/(T||H|-2|)$ and 
$C_2/(T||H|-2|^2)$, respectively.

To investigate the region in which Eqs.~(\ref{eq:T0}) and (\ref{eq:h0})
are applicable, and to examine whether there is any scaling behavior in the
prefactor near $|H|=2$ and $T=0$,
we performed Monte Carlo simulations with absorbing Markov chains 
(MCAMC)\cite{NOVO97,NOVO01} near $|H|=2$ at low temperatures, using the 
$d$-dimensional phonon dynamic [Eq.~(\ref{eq:wkl})]. 
Average lifetimes were measured over 2000 escapes with the linear system size 
$L=24$ and periodic boundary conditions. Since this system size is 
much larger than the critical droplet size near $|H|=2$, the dynamic behavior 
of the system does not depend on the examined system size. (This
was confirmed by comparing with simulations for $L=48$.) The range of 
temperatures examined was between $T=0.006$ and $T=0.10$, and the range of 
fields used was from $|H|-2=\pm 10^{-5}$ to $|H|-2=\pm 0.3$. 
At significantly low temperatures in a particular field or 
in a field quite close to $|H|=2$ at a particular temperature, 
the MPFUN package \cite{BAIL95} was used for arbitrarily 
high-precision calculations. Equation~(\ref{eq:lifetime}) was used
to obtain the prefactor from the measured average lifetimes. 

\begin{figure}
\begin{center}
\leavevmode
\epsfxsize=8.5cm
\epsfysize=7.5cm
\epsfbox{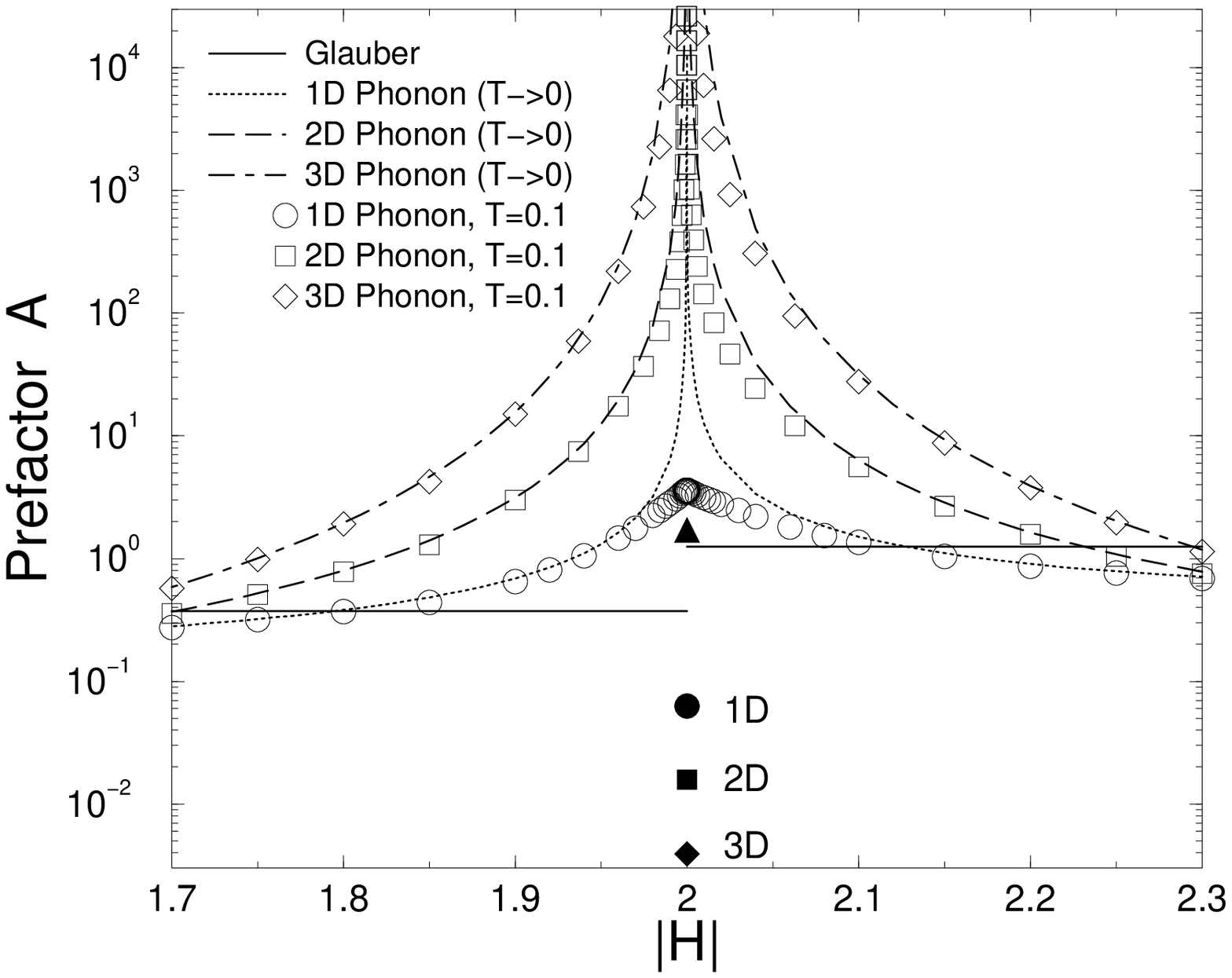}
\caption{The prefactor $A$ vs field $|H|$ for the Glauber dynamic
and the $d$-dimensional phonon dynamics. The open symbols are from the
Monte Carlo simulations at $T=0.1$ with the phonon dynamics. The curves are 
from the low-temperature analytic results Eqs.~(\ref{eq:a1}) and 
(\ref{eq:a2}) with the phonon dynamics, while the straight solid
lines are for the Glauber dynamic. For the $d=1$ simulation data 
with the phonon dynamic, a lower temperature or a field farther away 
from $|H|=2$ is needed to agree with the analytic results, compared to 
the simulation data for higher $d$. At $|H|=2$, the critical
droplet size changes, so the theoretical prediction Eq.~(\ref{eq:taulowT})
is not valid. The filled symbols are from the
Monte Carlo simulations with the phonon dynamics at $|H|=2$, except for
the filled triangle, which is from the Glauber dynamic. 
Here the prefactors at $|H|=2$ for both the phonon and Glauber dynamics were 
obtained by extrapolating low-temperature simulation data (from $T/J=0.04$ 
to $T/J=0.2$) to zero temperature.}
\label{fig:AvsH}
\end{center}
\end{figure}

Figure~\ref{fig:AvsH} illustrates the analytic and numerical (Monte Carlo
simulations at $T=0.1$) prefactors as functions of the field with the Glauber 
and the $d$-dimensional phonon dynamics. The analytic results for the
prefactor with the phonon dynamics diverge as $|H| \rightarrow 2^{\pm}$. 
This divergence does not occur with the Glauber dynamic 
(straight solid lines in Fig.~\ref{fig:AvsH}). 
The simulation data at $T=0.1$ agree well with the analytic results far away
from $|H|=2$, but they start to deviate from the analytic results as $|H|$ 
approaches 2. This deviation is more significant for lower-dimensional
phonon baths (smaller $d$) at a fixed temperature, as shown in 
Fig.~\ref{fig:AvsH}. For $d=3$ the simulation data at $T=0.1$ 
deviate from the analytic results only when the field is much closer to 
2 than for the $d=1$ simulation data. We also find that lower-dimensional 
phonon dynamics need much lower temperatures to agree with the analytic results
as $|H|$ approaches 2. From now on, we concentrate on the phonon dynamics
only. 

Figure~\ref{fig:logAvslogH} shows log-log plots of the prefactor as a 
function of field at different temperatures and for different $d$. 
For $d=1,2$, and 3, the simulation data show two regimes 
where Eqs.~(\ref{eq:T0}) and (\ref{eq:h0}) hold, respectively. Roughly speaking
(more precise statements will be made in Sec.V.), when the field is not 
too close to $|H|=2$ [about the right halves of 
Figs.~\ref{fig:logAvslogH} (a),~(b), and (c)],
below (or above) $|H|=2$, the prefactors at different temperatures converge
to a linear curve. For $d=1$, these converging linear curves
can be found for larger $||H|-2|$ at higher $T$. 
The slopes of the converging linear curves (this is easily 
seen for $d=2$ and $3$) are approximately $-d$, as expected from 
Eq.~(\ref{eq:T0}). When the field is close enough to $|H|=2$ 
(the left halves of Fig.~\ref{fig:logAvslogH}), the prefactor at 
fixed temperature again behaves linearly with $||H|-2|$ on a logarithmic scale.
As expected from Eq.~(\ref{eq:h0}), the slopes of those linear curves 
are approximately $-(d-1)$ and do not depend on temperature, but
the intercepts are different for different temperatures.

\section{Scaling}

Assuming that the prefactor $A$ is a generalized homogeneous 
function \cite{HANK72} of the field and temperature only, 
we can write it as
\begin{eqnarray}
A(\lambda^{a_h} h, \lambda^{a_T} T)&=&\lambda^{a_A} A(h,T) \;,
\end{eqnarray}
where $h=|H|-2$. Choosing $\lambda=T^{-1/a_T}$, this gives
\begin{eqnarray}
A(h, T)&=& \left(\frac{1}{T} \right)^{a_A/a_T} 
\Phi^{\pm}\!\left(\frac{h}{T^{a_h/a_T}}\right) \:, 
\label{eq:scaling}
\end{eqnarray}
where $\Phi^{\pm}(x)$ are scaling functions 
for $x=h/T^{a_h/a_T}>0$ and $x<0$, respectively. 
Since the prefactor has the asymptotic
behaviors shown in Eqs.~(\ref{eq:T0}) and (\ref{eq:h0}),
the scaling functions $\Phi^{\pm}(x)$ should have these properties:
\begin{equation}
\lim_{|x| \rightarrow \infty} \Phi^{\pm}(x) = \frac{1}{C_1 \: |x|^d} 
\: \: \:,\: \: \lim_{|x| \rightarrow 0} \Phi^{\pm}(x) = 
\frac{C_2}{|x|^{d-1}} \; .
\end{equation}
From the above properties, the scaling exponents are determined to be 
$a_A/a_T=d$ and $a_h/a_T=1$, so $A(h,T)=(1/T)^d \;
\Phi^{\pm}\! (h/T)$.

\begin{figure}
\begin{center}
\leavevmode
\epsfxsize=8.5cm
\epsfysize=7.5cm
\epsfbox{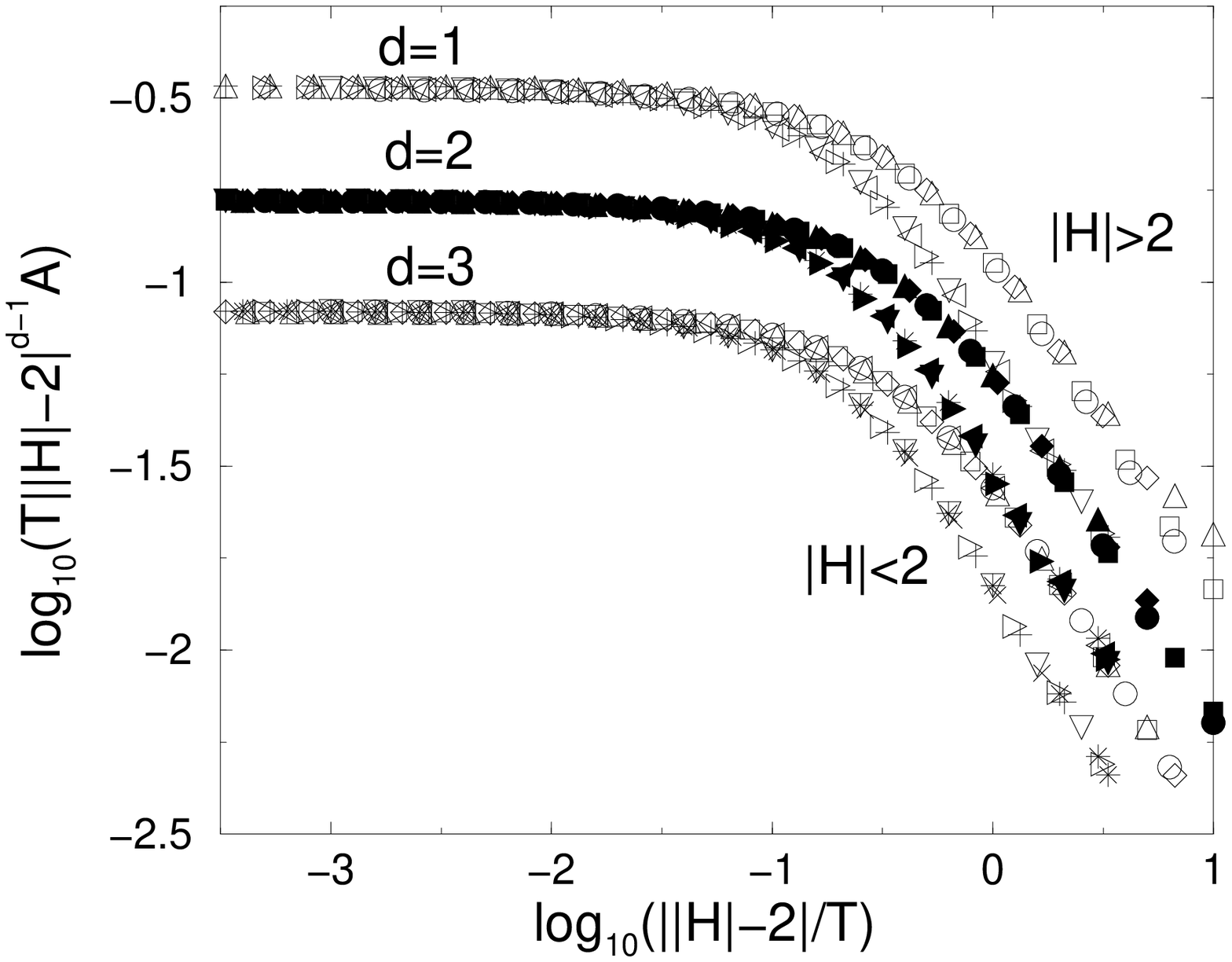}
\caption{Log-log plots of $T||H|-2|^{d-1} A(H,T,d)$ vs $x=||H|-2|/T$ for
different temperatures (for $d=1$, $T=0.006,0.01,0.02,0.03$; for $d=2$, 
$T=0.02,0.03,0.04,0.06$; and for $d=3$, $T=0.01,0.02,0.03,0.06,0.1$) and 
different-dimensional phonon dynamics. 
For each value of $d$, the upper curves
are for $|H|>2$, while the lower curves are for $|H|<2$. The simulation 
data for different temperatures collapse well for each value of $d$. 
Also, for large $x$, the curves for $d$ and $|H|<2$ coincide with
those for $d+1$ and $|H|>2$, and the curves multiplied by 2 for $d+1$
and $|H|>2$ ($|H|<2$) coincide with the curves for $d$ and 
$|H|>2$ ($|H|<2$).}
\label{fig:logAhTvslogH}
\end{center}
\end{figure}

\begin{figure}
\begin{center}
\leavevmode
\epsfxsize=8.5cm
\epsfysize=7.5cm
\epsfbox{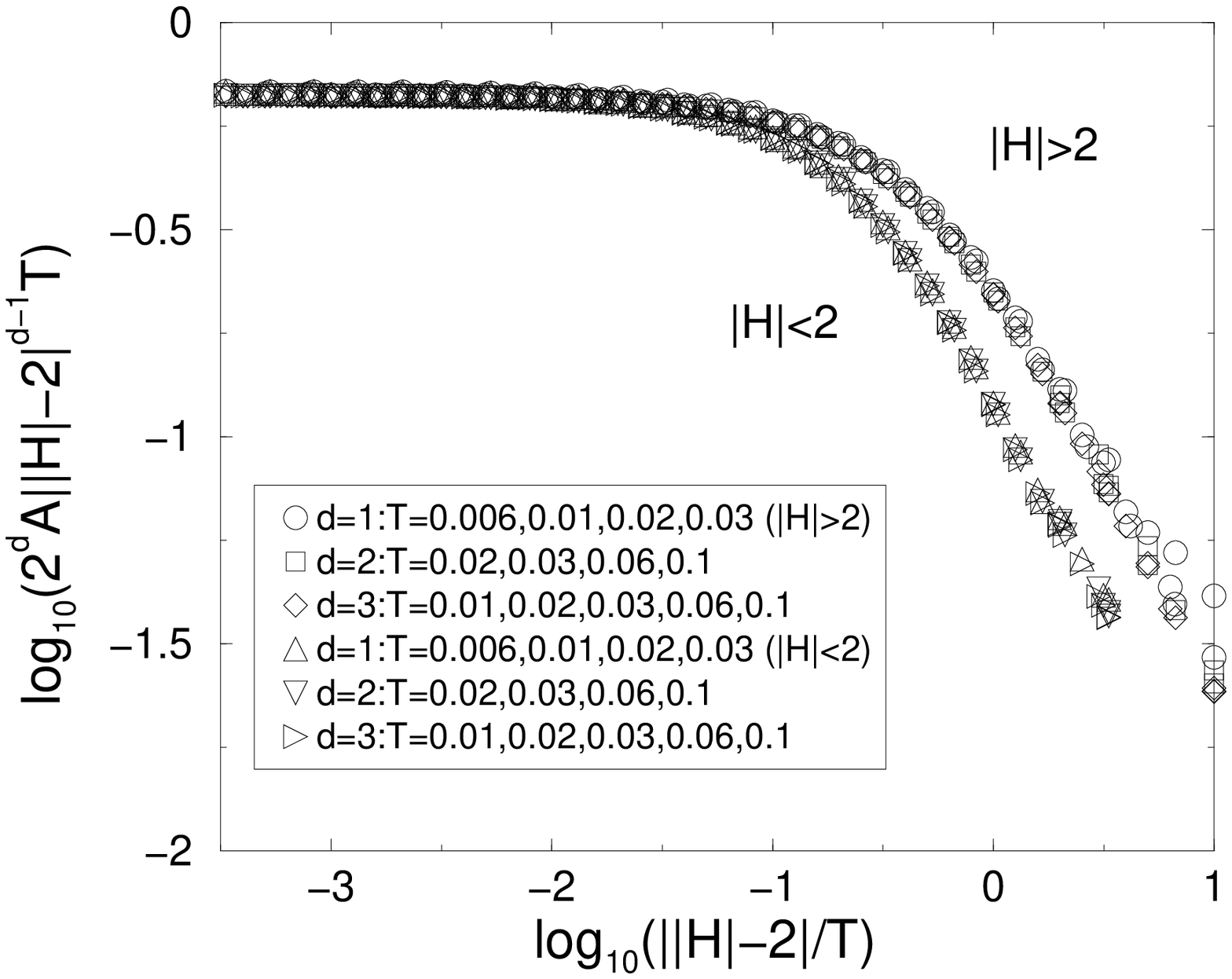}
\caption{Log-log plots of $T||H|-2|^{d-1} A(H,T,d)$ multiplied by $2^d$ 
vs $||H|-2|/T$ for different $T$ and $d$. For $|H|>2$ (or $|H|<2$), 
all the simulation data for different $T$ and $d$ collapse onto a single
curve. The scaling function
for $|H|>2$ is different from that for $|H|<2$. Here the saturation value of 
$2^d T||H|-2|^{d-1} A(H,T)$ as $||H|-2|/T \rightarrow 0$
(that is, $2^d \: C_2$) is approximately 2/3, so that 
$C_2 \approx 2/3 \cdot 1/2^d$.}
\label{fig:comb}
\end{center}
\end{figure}

If we rewrite the prefactor using scaling functions,
$\Psi^{\pm}(x)=|x|^{d-1}\Phi^{\pm}(x)$, which have these properties
\begin{equation}
\lim_{|x| \rightarrow \infty} \Psi^{\pm}(x) = \frac{1}{C_1 \: |x|} 
\; \; \; ,\; \; \; 
\lim_{|x| \rightarrow 0} \Psi^{\pm}(x) = C_2 \; ,
\label{eq:psi}
\end{equation}
then the prefactor can be written as
\begin{eqnarray}
A(h,T) &=& \frac{1}{T} \frac{1}{h^{d-1}} 
\Psi^{\pm}\!\left(\frac{h}{T}\right) \;.
\label{eq:scaling2}
\end{eqnarray}
Log-log plots of $T|h|^{d-1}A$ vs $|h|/T$ at different temperatures and for 
different $d$ are shown in Fig.~\ref{fig:logAhTvslogH}.
The simulation data for different temperatures collapse well onto
each other for each value of $d$ and for $|H|>2$ (or $|H|<2$).
Because $C_1$ for a particular $d$ and $|H|<2$ is the same as
$C_1$ for $d+1$ and $|H|>2$ [Eqs.~(\ref{eq:T0}) and (\ref{eq:psi})], 
the simulation data in the regime
of large $|h|/T$ for $d$ and $|H|<2$ fall on the data for $d+1$ and $|H|>2$.
Using the fact that for large $|h|/T$ and $|H|>2$ ($|H|<2$),
$C_1$ multiplied by 2 for $d$ is the same as $C_1$ for $d+1$, 
we make log-log plots of $T|h|^{d-1}A$ multiplied by $2^d$ vs $|h|/T$
for different temperatures and for different $d$ (see Fig.~\ref{fig:comb}). 
Then all the simulation data for different 
temperatures and different $d$ collapse well onto two master curves, 
one for $|H|>2$, and the other for $|H|<2$. 
The saturation value of $2^d T|h|^{d-1}A$ as $|h|/T \rightarrow 0$ 
($2^d \: C_2$) is approximately 2/3 (we do not know if this
value is exactly 2/3 or not), 
so that $C_2 \approx 2/3 \cdot 1/2^d$ and $C_2 \propto 1/C_1$. 
Using this approximate expression for 
$C_2$ and Eqs.~(\ref{eq:psi}) and (\ref{eq:scaling2}), 
we can confirm that the slopes of the log-log plots of 
$2^d T|h|^{d-1}A$ vs $x=|h|/T$ (Fig.~\ref{fig:comb}) for large $x$ 
are $-1$ for both $|H|>2$ and $|H|<2$, and that their intercepts at
$x=1$ for $|H|>2$ ($|H|<2$) are approximately $\log_{10}\: (1/4)$ 
($\log_{10}\: (1/8)$).

\section{Discussion and Conclusions}

\begin{figure}
\begin{center}
\leavevmode
\epsfxsize=8.5cm
\epsfysize=7.cm
\epsfbox{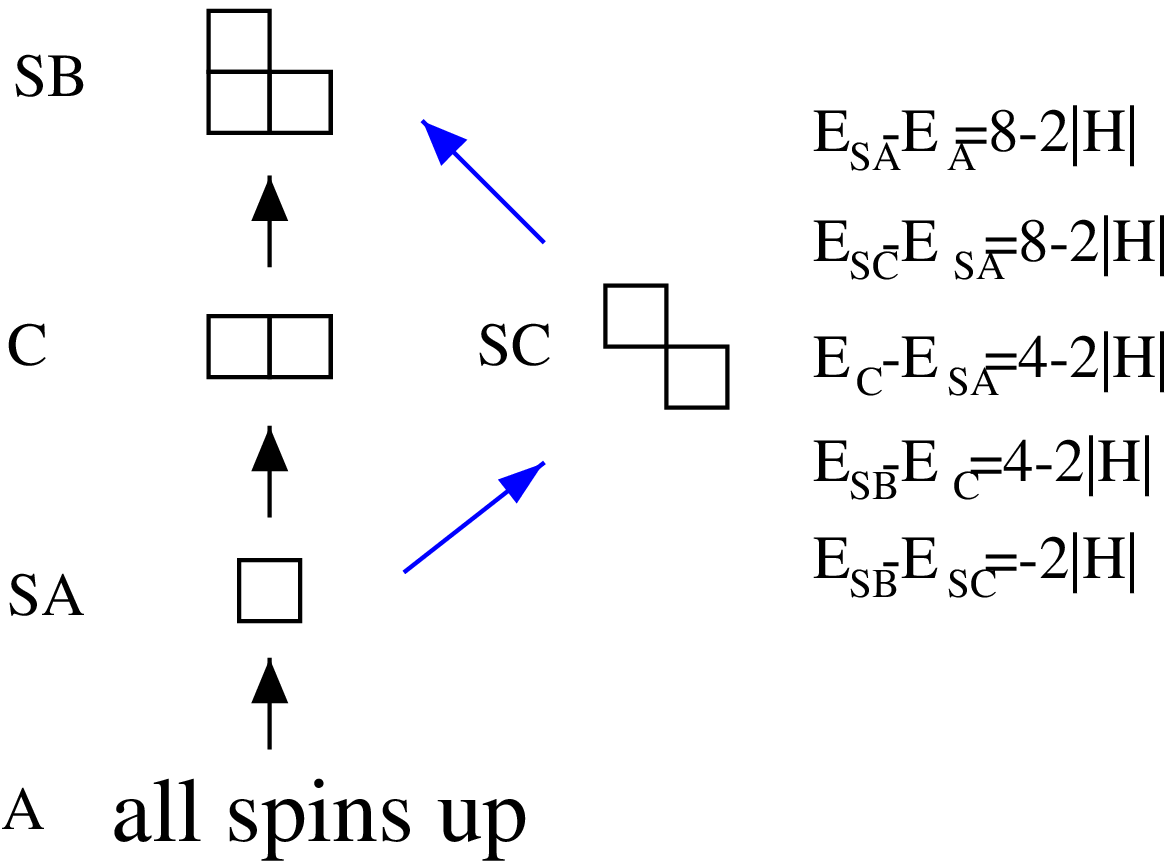}
\vspace{.3cm}
\caption{Schematic diagram that shows the most probable paths of the
system for different fields. Each box represents one overturned spin. Each
configuration is labeled. The configuration labeled SA is the
saddle point for $|H|>2$, SB for $|H|<2$, and SC for $|H|=2$.
The energy difference between two configurations is given on the 
right-hand side. As shown, the energy difference between SA and C 
is zero at $|H|=2$, and so is the
energy difference between C and SB. Therefore,
at $|H|=2$, the path from SA to SB through C is forbidden according to
the phonon dynamic, Eq.~(\ref{eq:wkl}). Consequently, the system chooses 
the saddle point SC to reach the configuration SB 
(A$\rightarrow$SA$\rightarrow$SC$\rightarrow$SB). 
Then the energy barrier at $|H|=2$ is 
$E_{\mathrm SC}-E_{\mathrm A}=2(8-2|H|)=8$ since the path
from SC to SB is downhill. }
\label{fig:conf}
\end{center}
\end{figure}

Our results indicate that the divergence of the lifetime
prefactor is caused by the phonon dynamic we imposed, not
by any unusual properties of the saddle points themselves, 
as is the case for Ref.~\cite{MAIE01}. 
Referring to Fig.~\ref{fig:conf}, we discuss saddle points and 
the most probable paths to escape from the metastable state 
for different fields. 

For $|H|>2$, the saddle point corresponds to the configuration SA,
only one overturned spin, so the most probable path to escape from
the metastable well is A$\rightarrow$SA.
As $|H| \rightarrow 2^{+}$, the energy difference between SA and C 
approaches zero (Fig.~\ref{fig:conf}), so the transition rate 
between them becomes zero. Consequently, the attempt frequency to grow 
the droplet along the downhill path SA$\rightarrow$C decreases. 
This is the reason that the lifetime prefactor diverges 
as $|H| \rightarrow 2^{+}$.
Then, a new probable path towards SC through SA begins to be preferred
to the downhill path. At $|H|=2$, the path SA$\rightarrow$C is
forbidden, and the most probable path is directed towards the
saddle point SC. 

For $|H|<2$, the saddle point corresponds to the configuration SB,
an L-shaped cluster of three overturned spins. For this field, the most 
probable path to escape from the metastable well is 
A$\rightarrow$SA$\rightarrow$C$\rightarrow$SB. As $|H| \rightarrow 2^{-}$, 
the energy difference between SA and C and the energy difference between 
C and SB both approach zero, so that the transition rates between them become 
zero. As a result, the attempt frequency along the most probable path 
towards the saddle point SB decreases, which leads to the prefactor
divergence as $|H| \rightarrow 2^{-}$. Then, a new probable 
path towards SC through SA develops (Fig.~\ref{fig:conf}). 
Thus, SC starts to compete with SB. At $|H|=2$, the attempt frequency 
towards SB is zero, and the most probable path is towards 
the saddle point SC.

\begin{figure}
\begin{center}
\leavevmode
\epsfxsize=5.5cm
\epsfysize=4.5cm
\epsfbox{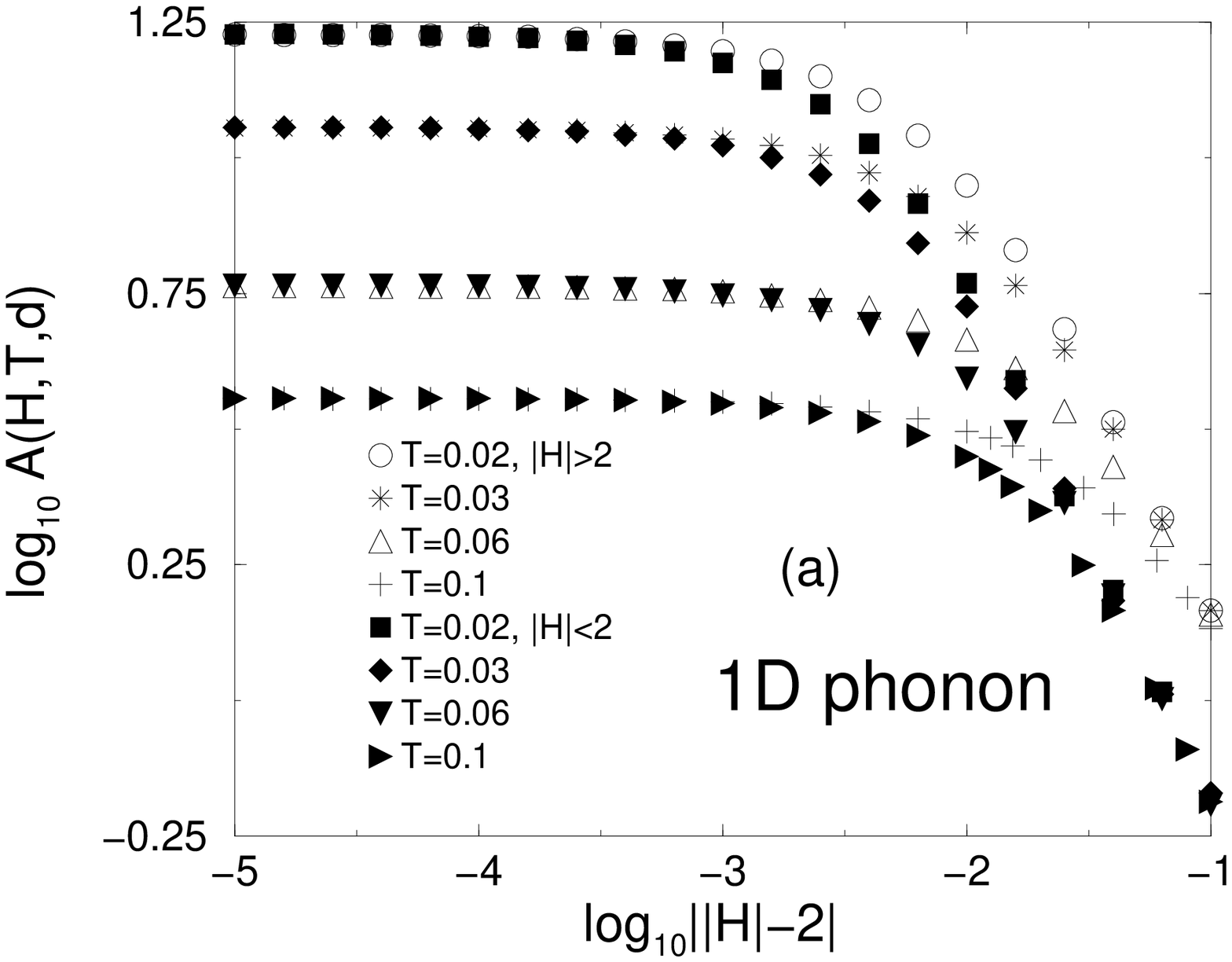}
\epsfxsize=5.5cm
\epsfysize=4.5cm
\epsfbox{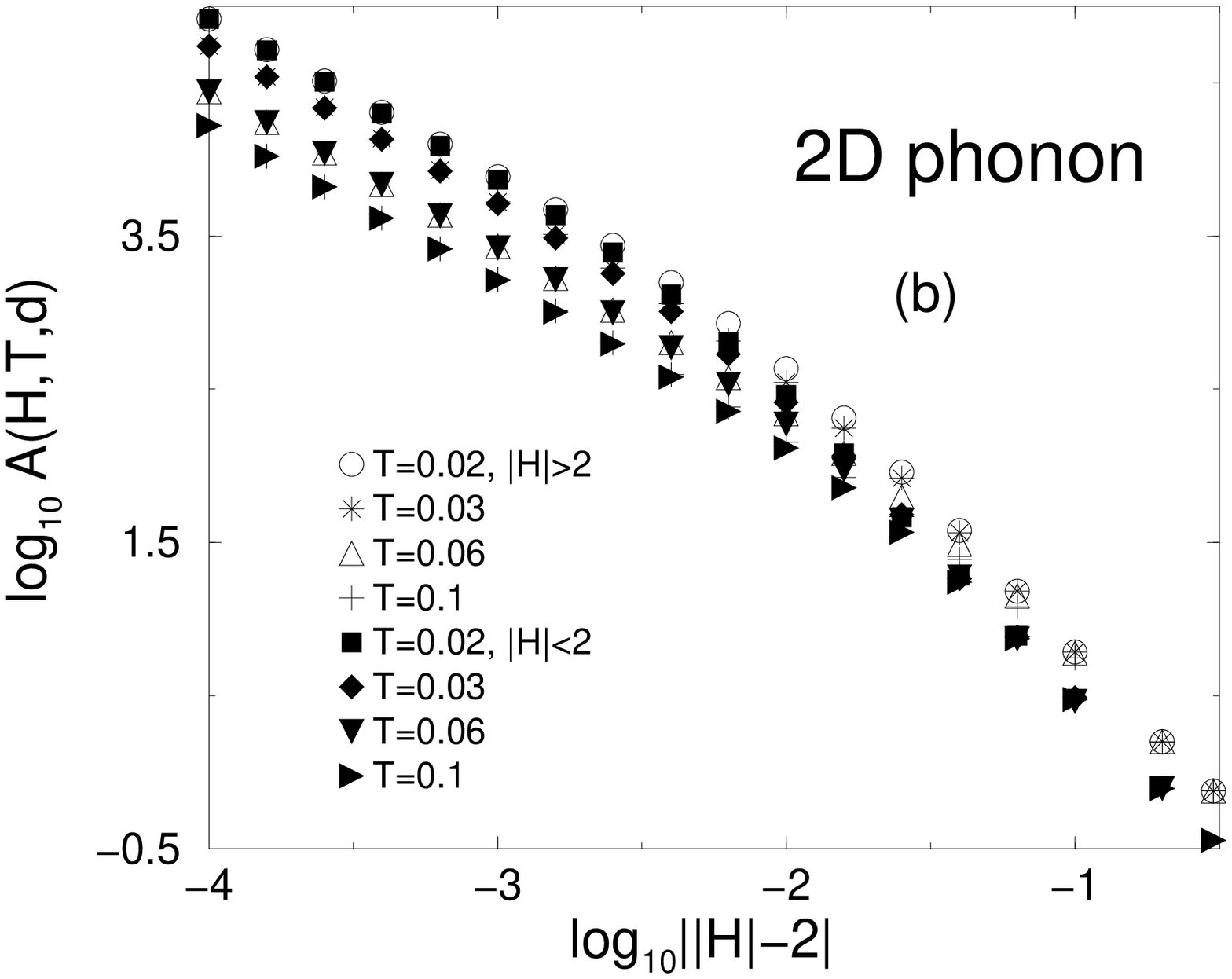}
\epsfxsize=5.5cm
\epsfysize=4.5cm
\epsfbox{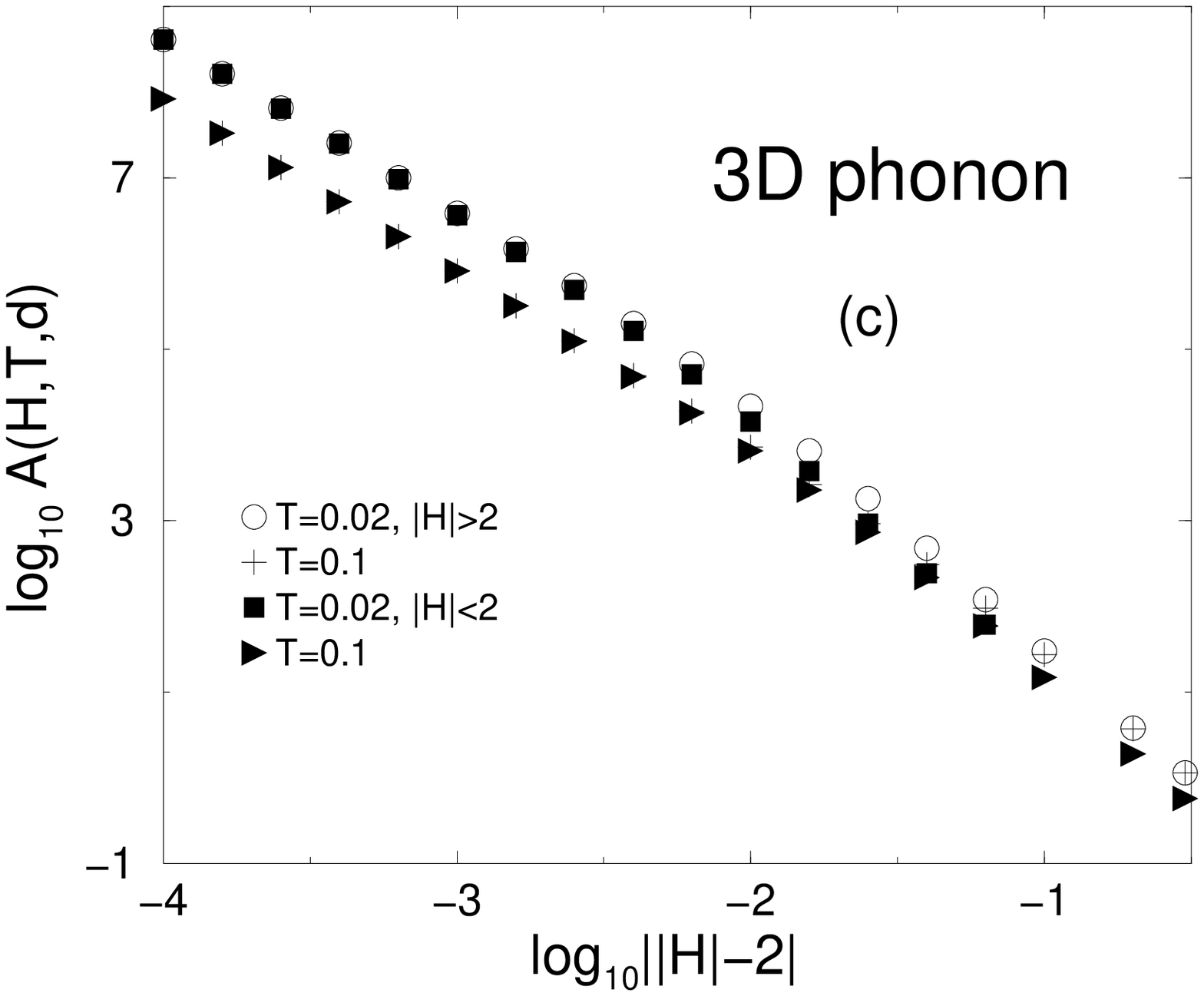}
\caption{Log-log plots of the prefactor $A(H,T,d)$ vs $||H|-2|$
at different temperatures for (a) $d=1$, (b) $d=2$, and (c) $d=3$ 
dimensional phonon dynamics, as the field $|H|$ approaches 2 from above
(empty symbols) and below (filled symbols). In (c), only data for two 
temperatures are shown for 
clarity, while the simulations were performed for $T=0.01,0.02,0.03,0.06$, 
and 0.1. There are two regimes in these plots: (1) As 
$||H|-2|/T \rightarrow 0$ (the left-hand side of the plots),
the slopes of the linear curves for different temperatures
become $-(d-1)$. (2) As $||H|-2|/T \rightarrow \infty$ (the
right-hand side of the plots), the curves for different temperatures
collapse onto single linear curves whose slopes are $-d$.
These results are predicted from Eqs.~(\ref{eq:T0}) and (\ref{eq:h0}).}
\label{fig:logAvslogH}
\end{center}
\end{figure}

The previously shown scaling behavior of the prefactor 
can be observed from the simulation data in Fig.~\ref{fig:logAvslogH}.
In the log-log plots of $A$ vs $||H|-2|$, a slope of $-(d-1)$ is found 
for the regime when $||H|-2|/T \rightarrow 0$, while a slope of $-d$ is 
found when $||H|-2|/T \rightarrow \infty$. The variable $||H|-2|/T$
(not simply either $||H|-2|$ or $T$) determines how those two regimes
change with field and temperature. The slope of $-(d-1)$ occurs at smaller
$||H|-2|$ at lower temperatures, in order to reduce $||H|-2|/T$,
compared to higher-temperature data. The slope of $-d$ occurs
at larger $||H|-2|$ at higher temperatures, in order to
increase $||H|-2|/T$, compared to lower-temperature data.

In summary, we have examined the prefactor of 
the lifetime of the metastable state
for a square-lattice Ising ferromagnet with a $d$-dimensional phonon bath
attached to each spin. Using the transition rates calculated from
the density-matrix equation, we demonstrated analytically and numerically
that the metastable lifetime prefactor diverges as $|H| \rightarrow 2$, 
and that it also scales as a function of $||H|-2|/T$ for each value of $d$, 
near $|H|=2$ and $T=0$. This scaling is a result of the fact that 
the prefactor is a generalized homogeneous function \cite{HANK72} 
of the field and temperature. The divergence and scaling of the prefactor 
are due to the chosen dynamic, not to any non-smooth energy landscape at the 
particular magnetic field.

\begin{center}
{\textbf{Acknowledgments}}
\end{center}

This work was funded by NSF Grant Nos.~DMR-9871455 and DMR-0120310, 
and by Florida State University through the School of 
Computational Science and Information Technology and
the Center for Materials Research and Technology.

%
%


\end{document}